 \documentclass[preprint2]{aastex}

\shorttitle{Imaging Polarimetry of Mrk 477}
\shortauthors{Kishimoto et al.}

\newcommand{\flam}{erg cm$^{-2}$ sec$^{-1}$ \AA$^{-1}$}
\newcommand{\crate}{cts sec$^{-1}$ pixel$^{-1}$}
\newcommand{\chisq}{$\chi^2$}
\newcommand{\ha}{H$\alpha$}
\newcommand{\hb}{H$\beta$}

\begin{document}

% ----- ----- ----- title, authors, abstract ----- ----- ----- -----

\title{UV Imaging Polarimetry of the peculiar Seyfert 2 galaxy Mrk
477}

\author{Makoto Kishimoto}
\affil{Physics Department, University of California, Santa Barbara, 
Santa Barbara, CA 93106}
\email{makoto@physics.ucsb.edu}

\author{Laura E. Kay}
\affil{Department of Physics and Astronomy, Barnard College, Columbia
University, New York, NY 10027}

\author{Robert Antonucci and Todd W. Hurt}
\affil{Physics Department, University of California, Santa Barbara, 
Santa Barbara, CA 93106}

\author{Ross D. Cohen}
\affil{Center for Astrophysics and Space Sciences, Code 0424, 9500
Gilman Drive, University of California at San Diego, La Jolla, CA
92093}

\and

\author{Julian H. Krolik}
\affil{Physics and Astronomy Department, Johns Hopkins University,
Baltimore, MD 21218}

\begin{abstract}

We present the results of UV imaging polarimetry of the Seyfert 2
galaxy Mrk 477 taken by the Faint Object Camera onboard the Hubble
Space Telescope (HST). From a previous HST UV image ($\lambda \sim
2180$\AA), Mrk 477 has been known to have a pointlike bright UV
hotspot in the central region, peculiar among nearby Seyfert 2
galaxies.  There are also claims of UV/optical variability, unusual
for a Seyfert 2 galaxy. Our data show that there is an off-nuclear
scattering region $\sim 0.''6$ ($\sim 500$ pc) NE from the hotspot.
The data, after the subtraction of the instrumental effect due to this
bright hotspot region, might indicate that the scattered light is also
detected in the central $0''.2$ radius region and is extended to a
very wide angle.  The hotspot location is consistent with the symmetry
center of the PA pattern, which represents the location of the hidden
nucleus, but our data do not provide a strong upper limit to the
distance between the symmetry center and the hotspot.  We have
obtained high spatial resolution color map of the continuum which
shows that the nuclear spiral arm of $0.''4$ scale ($\sim$ 300pc) is
significantly bluer than the off-nuclear mirror and the hotspot
region.  The nature of the hotspot is briefly discussed.

\end{abstract}


% ----- ----- ----- body of paper ----- ----- ----- -----

\section{Introduction}\label{sec-intro}

Mrk 477 is classified as a Seyfert 2 galaxy since its permitted
emission lines are narrow in the optical spectrum.  However, it has
some unusual properties among this class of objects. Seyfert 2
galaxies generally do not show variability in the UV/optical, but for
this object, some variations have been reported. De Robertis (1987)
first reported that the Fe lines ([Fe VII] $\lambda$6087 and [Fe X]
$\lambda$6375) are variable, and the optical continuum seems to have
increased by a factor of 2 between the 1980 and 1985 observations.
Although Veilleux (1988) pointed out that the reported variation in
the FeX line is due to the afterglow in the detector and the FeVII
line variation is questionable due to the presence of a sky line,
Kinney et al. (1991) also reported that the UV/optical continuum
increased by a factor of 2 over a $5-6$ year period, based on the
mismatch of the flux at $\sim 3000$\AA\ from the IUE data taken in
1984 (partly in 1983) and the optical spectra taken in 1989 (confirmed
in 1990).

Kinney et al. also suggested that its nuclear ionizing source does not
need to be blocked from direct view based on a photon budget argument.
This is strange in the sense that, if we are seeing the nucleus
directly, we should see broad permitted lines also directly, since the
nuclear continuum source is generally thought to be more compact than
the broad-line region. But in the Mrk 477 spectrum, we do not see the
broad lines in total flux (see discussion of upper limit in
\S\ref{sec-disc}).  This might lead us to consider an actual lack of a
broad-line region in this object. However, optical spectropolarimetry
has shown the presence of broad lines in the polarized flux spectrum
(Tran, Miller, \& Kay 1992; Tran 1995), suggesting that the broad-line
region does exist but is hidden from direct view and seen only through
scattered light. This is in accordance with the general idea for
Seyfert 2 galaxies that the continuum source and broad-line region
seen in Seyfert 1 galaxies exist also in Seyfert 2 galaxies but are
hidden from direct view (Antonucci \& Miller 1985).  This idea is
consistent with the fact that nearby Seyfert 2 galaxies generally do
not show an unresolved bright nuclear source in their HST UV/optical
image (Nelson et al. 1996; Malkan, Gorjian, \& Tam 1998). However, the
HST UV image (at $\sim 2180$\AA; Heckman et al. 1997) has shown that
Mrk477 has a fairly bright UV pointlike source in its central region,
which is again peculiar.  On the other hand, the HST/GHRS UV
spectroscopy has shown that there is strong starburst activity in a
rather compact nuclear region (a few hundred pc scale; Heckman et
al. 1997).

In order to reveal the detailed nuclear structure of this peculiar
Seyfert 2 galaxy, we have conducted HST imaging polarimetry of Mrk
477.  Our data have spatially resolved the nuclear polarization
structure.  We describe our observations in \S\ref{sec-obs} and the
results in \S\ref{sec-res}. The implication of the results are
discussed in \S\ref{sec-disc} and our conclusions are summarized in
\S\ref{sec-conc}. We adopt $H_0 = 65$ km sec$^{-1}$ Mpc$^{-1}$
throughout this paper. Mrk 477 is at $z=0.038$, so the distance is
$\sim$ 180 Mpc and one arcsec corresponds to $\sim$ 800 pc.  The
Galactic reddening for this object is low, $E(B-V) = 0.011$ (NED;
Schlegel et al. 1998). For the Galactic reddening correction, we adopt
the reddening curve of Cardelli et al. (1989) with $A_V/E(B-V) = 3.1$.

\section{Observation}\label{sec-obs}

Mrk 477 was observed on August 28 and 30, 1997 by the Faint Object
Camera (FOC) onboard the HST. The filters F275W ($\lambda \sim
2800$\AA) and F342W ($\lambda \sim 3400$\AA) were used with three
polarizers POL0, POL60, and POL120. All data were taken in the normal
$512 \times 512$ pixel mode, where the pixel size is $0.''014 \times
0.''014$ and the field of view is $\sim 7'' \times 7''$. The data are
summarized in Table~\ref{tab-data}. These were taken after the
installation of COSTAR.

The FOC has a complicated nonlinearity, which depends on the flux
distribution of the object in the frame. It has a 10\% level nonlinear
response for a count rate of 0.15 \crate\ (normal $512 \times 512$
pixel mode) for a uniform illumination. For a point source, the 10\%
nonlinearity occurs for a peak count rate of 1.0 \crate.  The central
region of Mrk 477 consists of a bright pointlike source surrounded by
a fainter extended structure.  Therefore, at the center of our image,
the instrument behavior will be at least between these two cases and
likely to be closer to a point source case.

In the F342W images, the recorded peak count rate at the central
pointlike source was $\sim 1.4$ cts sec$^{-1}$ pixel$^{-1}$.  This is
a little larger than the 10\% nonlinearity level for a point source
case.  In the extended structure surrounding the pointlike source, the
region with a count rate larger than the 10\% nonlinearity level for a
uniform illumination case (0.15 \crate) is of the size $\sim 10 \times
15$ pixel ($\sim 0.''14 \times 0.''22$).  To compare this size with
the wing of the point spread function (PSF; see section
\ref{sec-res}), we find that in the images of the PSFs scaled to have
peak counts of 1.4, the region with counts larger than 0.15 is of the
size $\sim 5 \times 6$ pixel. Thus, our images suffer from a small
nonlinearity.  We have applied a flat-field linearity correction (task
FFLINCOR in IRAF) on the raw images. The largest correction at the
peak location was about 30\%, and less than 10\% outside the $0.''1$
radius region. The nonlinearity would affect the polarization
measurement at least at the central $\sim 0.''2$ scale region, as we
will discuss in \S\ref{sec-res}.

For the F275W images, the recorded peak count rates were $\sim0.5$ cts
sec$^{-1}$ pixel$^{-1}$. The count rates were larger than 0.15 only
for the central $\sim 5 \times 5$ pixel region, which is also
essentially the region of the pointlike source, and the peak count
rate was well below the point source 10\% nonlinearity level.
Therefore, nonlinearity is expected to be small in these images. The
flat-field linearity correction was also applied to the F275W raw
images. The correction was less than 10\% even in the peak
region. However, the polarization measurement of the pointlike source
might be somewhat uncertain also in the F275W image, as discussed in
\S\ref{sec-res}.

After this nonlinearity correction, the data were processed in the
standard manner to correct for geometric distortion and flat-field
response. The reseau marks were removed using neighboring pixels. Each
set of the two exposures through each polarizer with F275W (see
Table~\ref{tab-data}) was checked for the telescope pointing by taking
a cross-correlation between the central 300 pixel $\times$ 300 pixel
regions of the two images. Only the set of POL120 images were
displaced slightly from each other, so we shifted one of the set
accordingly. After this, each set for F275W were coadded, resulting in
six images (two filters and three polarizers).

The three images with three polarizers, as well as the images through
different filters, are expected to be shifted relative to one another. 
The six images were registered using several pointlike sources in the
outer region of the images. The accuracy of this registration is
estimated to be 0.4 pixel, or $0.''006$, from the standard deviation
of these pointlike sources in the registered images.  Then the
background was subtracted using the outermost region of the images.

Each polarizer's throughput has a slightly different wavelength
dependence (especially POL60 in shorter wavelengths), so we need to
scale each polarizer's image accordingly. We have determined these
scaling factors by implementing synthetic photometry on the UV/optical
spectrum of Mrk 477, constructed from the FOS UV spectrum ($4.''3
\times 1.''4$ aperture; Cohen et al. 2002) and the ground-based
optical spectrum ($2''$ slit; Kay 1994), using the throughput for each
combination of three polarizers and two filters.  Then, each set of
the three images with three polarizers are combined to produce the
Stokes $I, Q, U$ images.

Using the same spectrum, we also estimated the narrow emission line
contamination to be $\sim 15$\% in the F342W filter and $\sim 10$\% in
the F275W filter.  For small spatial bins, the line contamination
could be different from this estimation. Since the narrow lines are
unpolarized, the line contamination simply dilutes $P$.  It
essentially will not affect the $Q$ and $U$ (unnormalized Stokes
parameters), though the spatial variation of the effective
transmittance would slightly affect them: the resulting uncertainty in
$P$ is estimated to be less than $\sim 1$\% in the F275W filter, and
much smaller in the F342W filter where the polarizer transmittances
are much similar to one another.

\section{Results}\label{sec-res}

\subsection{Overall description}

Figures \ref{fig_f342_papol} and \ref{fig_f275_papol} show the near-UV
(F342W) and UV (F275W) polarization images of the central $\sim 2''$
region of Mrk 477. The polarization $P$ has been calculated with 10
pixel ($0.''14$) $\times$ 10 pixel bins, and the regions with
statistical S/N in $P$ larger than 3 are shown with the lines at each
point proportional to $P$.  The position angle (PA) of polarization is
approximately centrosymmetric in the region $\sim 0.''6$ ($\sim 500$
pc) NE from the bright pointlike source (note that the north is not to
the top), indicating that this region is scattering the light from a
compact source. We call this region the NE mirror hereafter in this
paper.

However, in other regions, the polarization pattern is very
unusual. At about $0.''5$ west from the hotspot, there is a region
with a ``radial'' pattern of large polarization degrees. We call this
a large-P radial feature, for convenience in this paper. This is
primarily due to the radial-shape counts in the POL120 polarizer
images at this position (counts at this position are elongated along
the radial direction with respect to the brightest center) where the
count rate is $\sim 0.2$\% of the peak count rate. A similar but
weaker feature is also seen in the POL60 images.  This feature is seen
in both of the F342W and F275W (but weaker) filter images, so it is
not considered to be due to incidental high background counts at this
position.

The central $0.''4$ scale region consists of a few distinct features,
as discussed by Heckman et al. (1997) : (1) a bright pointlike source,
which we call here a hotspot, (2) $\sim 0.''2$ scale fuzz (or broken
ring feature as Heckman et al. called it) elongated roughly
north-south around the hotspot, (3) a fainter spiral arm of $\sim
0.''4$ scale. This central region also has a somewhat radial-like PA
pattern. We show slightly zoomed polarization maps in Figures
\ref{fig_f342_papf} and \ref{fig_f275_papf}, where the polarized flux
in each bin is drawn and the regions of statistical S/N in $P$ down to
1 are included to show the detailed polarization measurement at this
central region. In the following, we infer that the radial-like PA
pattern in the central $0.''4$ region is primarily due to an
instrumental effect from the compact bright hotspot + fuzz feature,
and also infer that the large-P radial feature is probably due to the
same effect.

\subsection{Instrumental polarization pattern}

\begin{figure}
\plotone{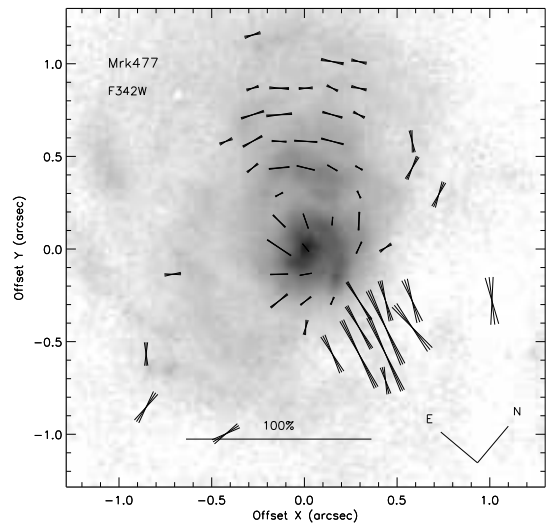}
\figcaption{Polarization map of Mrk 477 with the F342W filter. The
polarization $P$ is calculated with 10 pixel ($\sim 0.''14$) bins, and
the regions with S/N in $P$ larger than 3 are shown. The lines at each
point are proportional to $P$, and $1''$ length corresponds to
100\%. The three lines at each point indicates $1\sigma$ statistical
error of the PA measurement ($\theta_{\rm PA} \pm \sigma_{\theta_{\rm
PA}}$). The grayscale image is the total flux image through the F342W
filter in log scale. \label{fig_f342_papol}}
\end{figure}

\begin{figure}
\plotone{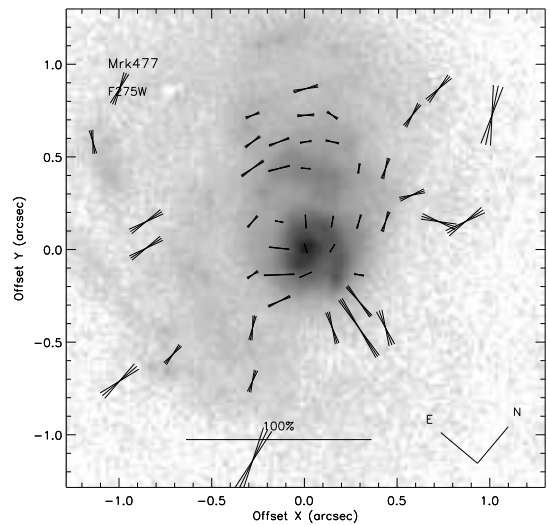}
\figcaption{The same as Fig.\ref{fig_f342_papol} but with the
F275W filter. The grayscale image is the total flux image through the
F275W filter in log scale. \label{fig_f275_papol}} 
\end{figure}

\begin{figure}
\plotone{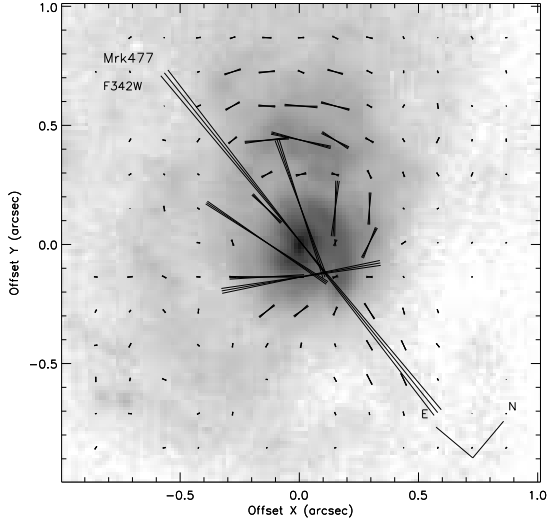}
\figcaption{The same as Fig.\ref{fig_f342_papol}, but slightly zoomed,
and the lines are drawn proportional to the polarized flux, where
$1''$ length corresponds to $1\times10^{-17}$ erg cm$^2$ sec$^{-1}$
\AA$^{-1}$. The regions with statistical S/N in $P$ larger than down
to 1 are shown.
\label{fig_f342_papf}} 
\end{figure}

\begin{figure}
\plotone{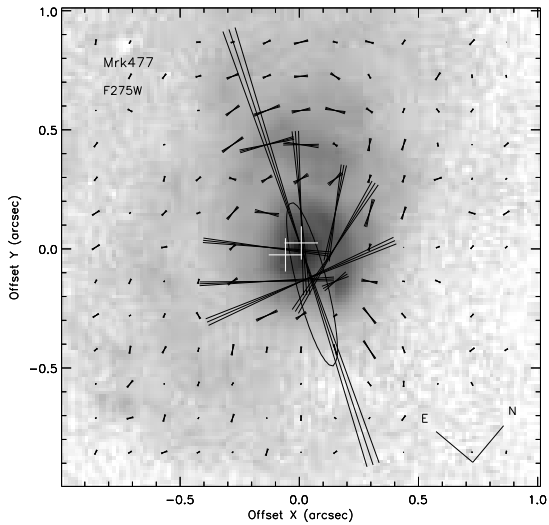}
\figcaption{The same as Fig.\ref{fig_f342_papf} but with the F275W
filter. The position of the symmetrical center of the PA pattern seen
in the NE mirror region is indicated as plus signs. The one on the
right side is for the F275W data and the left for the F342W data. The
error circle of 99\% confidence level is shown as a contour. See
\S\ref{sec-disc-nucpos} for the details.
\label{fig_f275_papf}} 
\end{figure}

The FOC polarimetry is implemented by taking images through three
different polarizers, called POL0, POL60, and POL120, which have axes
of three different directions with respect to the detector, 0\degr (or
180\degr), 60\degr (or 240\degr), and 120\degr (or 300\degr)
counter-clockwise from the image $+x$-direction. Figures
\ref{fig_f342_papol}$\sim$\ref{fig_f275_papf} are all preserved to
have the detector coordinate direction (so the north is {\it not} to
the top). The images through the three polarizers are known to have
different point spread functions (PSFs), which results in yielding
``instrumental polarization'' when the polarization of a compact
source is measured with small apertures (Hodge 1995). In fact, this
difference in the PSFs would also yield some instrumental features
when polarization is calculated at the regions around a bright
source. This instrumental effect has not been discussed in the
literature\footnotemark[1]. In order to assess this effect rather
quantitatively, we have constructed high S/N PSFs for each polarizer
using the data from the calibration observation of an unpolarized
globular cluster NGC 5272 (PID 5522; see Hodge 1996). We have added up
the images of a few of the brightest stars from the data sets taken in
two different epochs (1994 Aug 4 and 1995 May 23).

\footnotetext[1]{M. Ouchi, M. Kishimoto, \& R. Antonucci have noticed
this effect in the HST/FOC data of 3C109 (PID 6927) and the
calibration data (PID 6197), based on a suggestion by
R. Jedrzejewski.}

\begin{figure}
\plotone{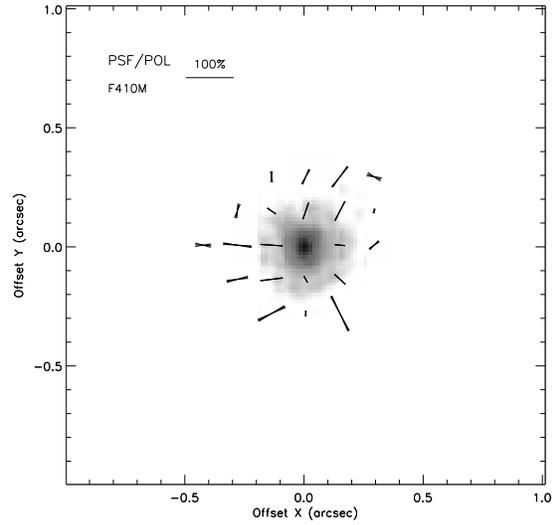}
\figcaption{The position angle map of the FOC polarization calibration
data. This is a sum of the observations of 5 unpolarized stars (3 and
2 in each of the two different epochs) taken with the F410M
filter. The 10 pixel bins with statistical S/N in $P$ larger than 1
are shown. The direction of the images are preserved to be in that of
the detector. The lines are proportional to $P$, where $0.''2$ length
corresponds to 100\%. The grayscale image is the total intensity image
in log scale. Note that the size of Figures \ref{fig_f342_papf}
$\sim$ \ref{fig_f275_papf_sub} is the same.
\label{fig_psf_papol}}
\end{figure}

\begin{figure}
\plotone{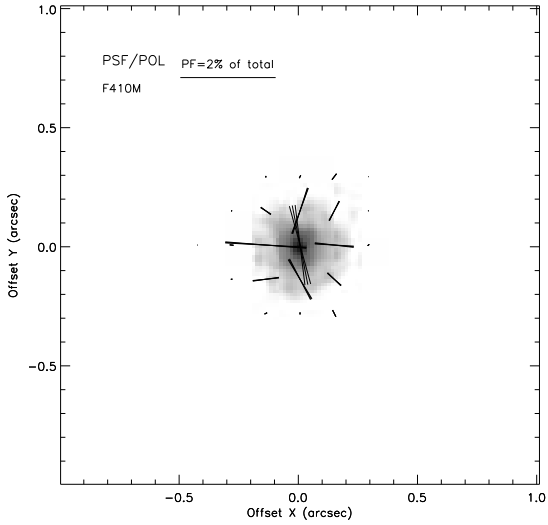}
\figcaption{The same as Fig.\ref{fig_psf_papol}, but the lines are
proportional to polarized flux in each bin. The $0.''4$ length
corresponds to 2\% of the total flux of the PSF.
\label{fig_psf_papf}}
\end{figure}

The PSFs through POL60 and POL120 were found to be slightly elongated
along the polarizer axis (for images of one star, see Fig.1 in Hodge
1995), though for POL60 the structure in the wing appears to be more
complicated.  This essentially seems to result in having an
instrumental radial PA pattern in the PSF wings. This is shown in
Figures \ref{fig_psf_papol} and \ref{fig_psf_papf} with the
``instrumental polarization'' and ``instrumental polarized flux'',
respectively, when the polarizations are calculated at the PSF wings
(this is an artifact that emerges from calculation, and not the
physical polarization). The instrumental polarized flux is very small
compared to the total flux of the source. Figure \ref{fig_psf_papf}
shows this ratio when binned with 10 pixels ($0.''14 \times
0.''14$). The ratio is at about the $1-2$ \% level at the bins
adjacent to the peak bin.  When the source is more extended and the
intensity gradient is much smaller, this instrumental effect will be
more smeared and become smaller.

For Mrk 477, however, the intensity gradient around the hotspot + fuzz
seems to be large enough that we see this effect. The PA pattern seen
in the central region (Figs. \ref{fig_f342_papol} and
\ref{fig_f275_papol}) actually resembles the PA pattern in the PSF
wings (Fig. \ref{fig_psf_papol}). Note that they are all in the same
detector coordinate directions and with the same size bins, so we can
directly compare them. These features in the PSF wings would depend on
the telescope focus/breathing, but the overall correspondence suggests
that our assessment of the PSF effect is at least approximately
correct.

In Figure \ref{fig_psf_papol}, we see a radial polarization feature
with a large $P$ in the lower-right direction from the center. This is
essentially due to the PSF wing through the POL120 polarizer.  This
feature seems to be similar to the large-P radial feature seen in our
Mrk 477 images, though not at the same distance from the central
source but almost in the same direction.  This could suggest that the
large-P feature seen in the Mrk 477 images would possibly be from the
POL120 PSF wing of the bright hotspot+fuzz with much larger counts.
The feature in our Mrk 477 images emerged even more clearly, perhaps
because it landed accidentally in the low count region which is still
rather close to the bright central region (see the overall diffuse
structure in Fig.\ref{fig_f342_papol} $\sim$ \ref{fig_f275_papf}).
The feature in the Mrk 477 images has a large $P$ but a small
polarized flux (see Figs. \ref{fig_f342_papf} and
\ref{fig_f275_papf}), and this is consistent with the small polarized
flux at the lower-right feature seen in the PSF data
(Fig.\ref{fig_psf_papf}).

One data set of FOC imaging polarimetry which might have been affected
by this instrumental effect is the one for NGC 1068 (Capetti et
al. 1995; Kishimoto 1999), which has a fairly bright and rather
compact source at the central region. In the detailed re-analysis of
this data set by Kishimoto (1999), the regions with a large intensity
gradient have been masked out, so that the analysis is not affected by
this effect.  In the FOC imaging polarimetry data of Mrk 3 (Kishimoto
et al. 2002), the images consist of multiple resolved clumps with much
lower counts compared to the peak in the Mrk 477 image, and the images
are not dominated by a single compact region. Therefore the data set
of Mrk 3 is essentially not affected by the instrumental effect
discussed above.

\subsection{Subtraction of the instrumental effect}

In addition to the rough similarity between the observed PA pattern
around the hotspot and that in the PSF wings, there appear to be some
possible systematic differences.  Therefore, we have attempted to
subtract the instrumental effect of the bright compact components,
i.e. hotspot + fuzz, using these PSF images. These two components are
compact and bright enough to have a steep intensity gradient which
produces the instrumental polarization discussed above.  The hotspot +
fuzz can be approximately modeled by a PSF + an elliptical gaussian
with its major axis at $\sim$ 25\degr\ clockwise from the image $+y$
direction (Fig.\ref{fig_f342_papol}; PA $\sim$ 15\degr\ in the sky
coordinates; see next section for the hotspot size): we have
subtracted out each polarizer's PSF from each polarizer's image of Mrk
477 by estimating the counts in the hotspot using synthetic aperture
measurement, and this left roughly an elliptical gaussian distribution
for the fuzz. We measured the FWHM of the elliptical gaussian in its
major and minor axis and subtracted in quadrature the FWHM of each
PSF, which resulted in $\sim$ 10 and 5 pixels, respectively. Then this
elliptical gaussian shape is convolved with the PSF through each
polarizer, and added with the PSF for the hotspot. The intensity ratio
of these two components can be roughly determined from the counts
ratio in the image in each of the F342W and F275W images (the
intensity ratio of the hotspot to the fuzz was found to be
approximately 1/3 and 1/4 in F342W and F275W, respectively).  Finally,
we have subtracted this two central features from each polarizer image
of Mrk 477, and reconstructed the polarization map.

\begin{figure}
\plotone{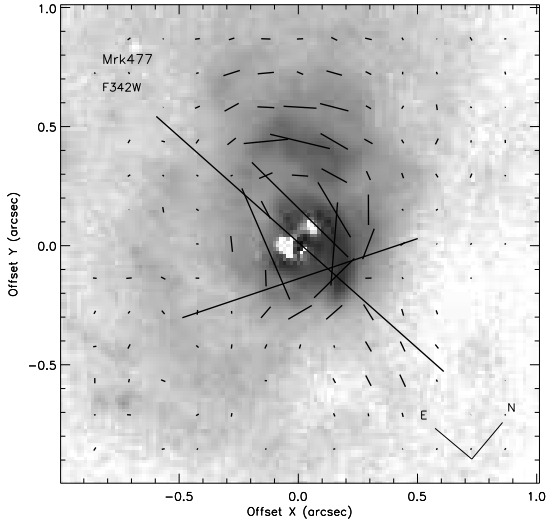}
\figcaption{The polarization map for the F342W filter after the
central compact components of $\sim 0.''2$ scale (hotspot + fuzz) has
been subtracted from each polarizer image, using the PSF through each
polarizer. See text for details on the subtraction process.  The lines
at each bin are drawn proportional to the polarized flux, where $1''$
length corresponds to $1\times10^{-17}$ erg cm$^2$ sec$^{-1}$
\AA$^{-1}$ (the same as in Fig.\ref{fig_f342_papf}).
\label{fig_f342_papf_sub}}
\end{figure}

\begin{figure}
\plotone{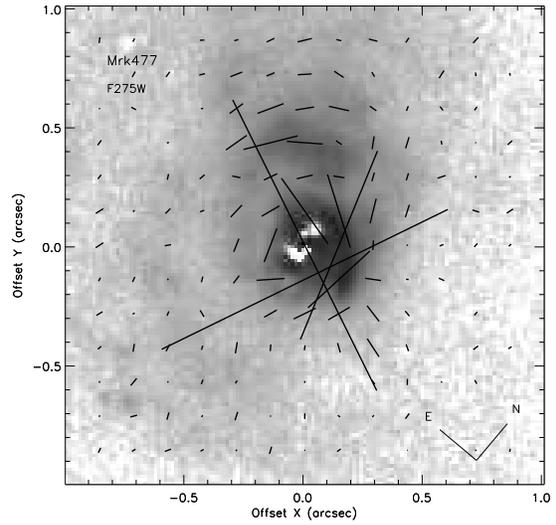}
\figcaption{The same as in Fig.\ref{fig_f342_papf_sub}, but for the
F275W filter.
\label{fig_f275_papf_sub}}
\end{figure}

The polarization images after this subtraction process are shown in
Figures \ref{fig_f342_papf_sub} and \ref{fig_f275_papf_sub}. The
radial PA pattern seems to be almost gone, except for the large-P
feature at the lower-right side of the image, and the PA pattern now
appears rather centrosymmetric around the hotspot.  Our modeling of
the central intensity distribution is very rough, but these results
are not so sensitive to the adopted parameters. We have confirmed this
by changing the parameters in a reasonable range.  For given PSFs, the
distribution of the instrumental polarized flux is roughly determined
by the approximate size and flux of the compact sources, and not by
the detailed shape of the sources.

However, we do not intend to claim that this is a definite subtraction
result. One large uncertainty in the subtraction is that the PSF wing
features would depend on the telescope focus/breathing at the time of
the observation. Therefore we also tried to use other PSF data from
the polarization calibration data (PID 6197, taken on 1996 July 12,
with the F342W filter). We have obtained a very similar PA pattern
from the NE to SW. The features at the east to south side of the
hotspot, however, were rather hard to subtract out with this different
set of PSFs.  Also, in the subtraction with the previous set of PSFs
(Figs \ref{fig_f342_papf_sub} and \ref{fig_f275_papf_sub}), there
might be over/under subtraction of the instrumental polarized flux in
this east to south region, since the polarization pattern after the
subtraction is different in the different filter image.  Thus the
polarization at this side is particularly uncertain in our analysis.

We note that in this process we essentially have subtracted
unpolarized components for the hotspot + fuzz.  The polarization of
this central region seems to be rather low, as discussed in the next
section.

\subsection{Polarization of the hotspot and NE mirror}

We estimate the size of the hotspot in our image to be FWHM $\sim$ 2.8
pixel ($0.''040$, $\sim$ 30pc) in the F342W image (from POL0 and
POL120 images; the core of the POL60 PSF is slightly degraded), from
the images with the fuzz feature roughly subtracted using an
elliptical gaussian (see previous section). This is almost consistent
with the PSF size through polarizer with the F342W filter, which is of
FWHM $\sim$ 2.6 pixel (note that we have primarily used the F410M
filter PSF of much larger counts, which has FWHM of $\sim$ 2.8 pixel).

As mentioned above, in the subtraction process we essentially
subtracted unpolarized components for the hotspot and fuzz, and we
have roughly fitted these unpolarized components in all the three
polarizer images assuming low polarization for the hotspot and
fuzz. We now turn our attention to the polarization of the hotspot.
The formal measurement with a synthetic aperture of 10 pixel diameter
($0.''14$) gives the polarization of the hotspot (plus some portion of
the fuzz) $P$ = 5.5\% at PA = 81\degr\ for F342W with statistical
error of $\sigma_P = 0.2$\% and $\sigma_{\rm PA} = 1.3$\degr\ (total
flux $F_{\lambda} = 3.0\times 10^{-16}$ \flam, corrected for the
Galactic reddening), 5.4\% at 61\degr\ for F275W with statistical
error of $\sigma_P = 0.3$\% and $\sigma_{\rm PA} = 1.6$\degr\
($F_{\lambda} = 3.4 \times 10^{-16}$ \flam). These polarization
measurements are slightly larger than the uncertainty from the PSF
differences : the same synthetic aperture polarimetry on the PSF shown
in Figure \ref{fig_psf_papol} yields $P$ = 2.7\%.  The POL120 images
of Mrk 477 has slightly larger count rate at the center than in the
POL0 and POL60 images, so the PAs obtained above for the hotspot is
approximately along the POL120 polarizer axis which is at PA =
70.05\degr (note that the image $+x$ direction was at PA =
$-49.95$\degr\ at the time of the observation of Mrk 477).  Therefore,
we infer that the hotspot is probably polarized at least a few percent
level at PA close to the POL120 axis.  This would be consistent with
the fact that the diagonal pattern noise which is often associated
with large count rate portion is slightly seen only in the POL120
image.

One uncertainty is the effect of nonlinearity at the hotspot location
on the polarization measurement. However, if the nonlinearity is mild
(we expect this to be the case; see \S\ref{sec-obs}) and the object is
not highly polarized, the nonlinear behavior of the detector would be
almost the same for the three images with three polarizers except that
the nonlinearity would make polarization detection smaller since the
difference in the counts in the three polarizers would become slightly
smaller.  As described in \S\ref{sec-obs}, we have attempted to
correct each image for nonlinearity by using FFLINCOR in IRAF. The $P$
measurement using the same synthetic aperture above on the data
without this correction yielded slightly smaller $P$ as expected
($P=4.5$\% at PA=80\degr\ with F342W, $P=5.0$\% at PA=59\degr\ with
F275W).

We can attempt to check the polarization measurement at the central
region for the F342W filter based on two measurements : $P$
measurement at the NE mirror and the ground-based optical $P$
measurement, which are summarized in Table \ref{tab-comp}.  For the
error calculation, we assumed a 5\% uncertainty in the background
counts estimation and added this in quadrature to the statistical
error.  We have measured the polarization of the NE mirror with a
synthetic aperture of $0.''8 \times 0.''6$ centered at $(0.''00,
0.''55)$ in Figure \ref{fig_f342_papf}, which yielded the PA of the NE
mirror to be 125\degr\ (Table \ref{tab-comp}). For the ground-based
measurement, we have used the data of Kay (1994) taken with a $2''$
slit and integrated over $3200-3600$\AA. The integrated PA is
calculated to be 100\degr\ (Table \ref{tab-comp}) and this is also
consistent with the the PA at longer wavelength : the PA in the
optical wavelength is also about 100\degr\ (Kay 1994; Tran 1995). This
difference between the PA of the NE mirror and that of the whole
central region suggests that there is another polarized component in
the region of hotspot + fuzz + arm. The polarization measured above in
the F342W filter for the hotspot plus some portion of the fuzz is
consistent with being this residual component.  A large-aperture
polarization measurement with the F342W filter is apparently
consistent with the ground-based measurement, despite that the FOC
measurement at this low polarization level could be affected by other
systematic errors.

However, the pre-COSTAR FOS measurement, taken with a $4.''3 \times
1.''4$ aperture with the minor axis at PA = $-124$\degr\ (Cohen et
al. 2002), integrated over $2500-3100$\AA, gives a different PA from
that with a synthetic aperture measurement simulated on the FOC F275W
filter data, as quoted in Table \ref{tab-comp} (the quoted error for
this large synthetic aperture FOC measurement is essentially from the
background subtraction uncertainty).  On the other hand, the PA at the
NE mirror from the FOC F275W filter data is similar to the
large-aperture FOS measurement (the synthetic aperture misses only a
minor portion of the NE mirror). However, the polarized flux from the
NE mirror seems to be significantly lower than that detected in the
FOS aperture.  This polarized flux difference suggests either that
there would be significant polarized flux in the hotspot + fuzz + arm
region at almost the same PA as that of the NE mirror or that the FOS
measurement suffers from some systematic error or both of these. In
the former case, there would be some systematic error in the central
region of the FOC F275W filter measurement and some possible effect
from nonlinearity.

The measured PA of the NE mirror with the synthetic aperture is
different in the F275W and F342W filters.  If we compare the PA
distribution within the aperture, we find that the F342W PA is
systematically rotated clockwise especially in the south-eastern
portion of the NE mirror.  This might be the effect of the
instrumental polarized flux, since this portion is symmetrically
opposite side of the large-P radial feature with respect to the
central hotspot region. If this is the case, the true polarized flux
from the NE mirror could be slightly larger than measured, since the
radial polarized flux would cancel the centrosymmetric polarized flux.
We estimate that the polarized flux might be larger than the value
quoted in Table \ref{tab-comp} by up to $\sim$ 30\% in both of the
filters, by measuring the instrumental radial polarized flux at the
large-P radial feature and also by using the synthetic aperture
symmetrically opposite to the aperture for the NE mirror. Even if we
incorporate this possible effect, the above argument is not
significantly affected.

\subsection{Color of polarized flux and total flux}\label{sec-res-pfcolor}

The polarized flux measurements in the two filters provide color
information. The limited S/N in our images and the instrumental effect
in the central region did not provide a reliable detailed map of the
polarized flux color, but we can obtain the color integrated over the
NE mirror region.  From the large synthetic aperture data for the NE
mirror (Table \ref{tab-comp}), the overall polarized flux color is
calculated to be $\alpha = -3.0 \pm 1.0$, where $P F_{\nu} \propto
\nu^{\alpha}$ (corrected for Galactic reddening).  There might be an
additional uncertainty from the possible effect of the instrumental
polarized flux, but the contamination in both of the two filters would
tend to cancel out.  This color is redder than the typical spectral
index of Seyfert 1 galaxies measured in the two filters F275W and
F342W, which is about $-0.9$ (Kishimoto et al. 2002). Note that the
color from these two filters would be slightly different from the
color of the true polarized continuum or true Seyfert 1 continuum,
since the two filters are on the so-called 3000\AA\ bump. The value of
$\sim -0.9$ as a typical Seyfert 1 color has been obtained
specifically for the two filters (see Kishimoto et al. 2002 for
details).

\begin{figure}
\plotone{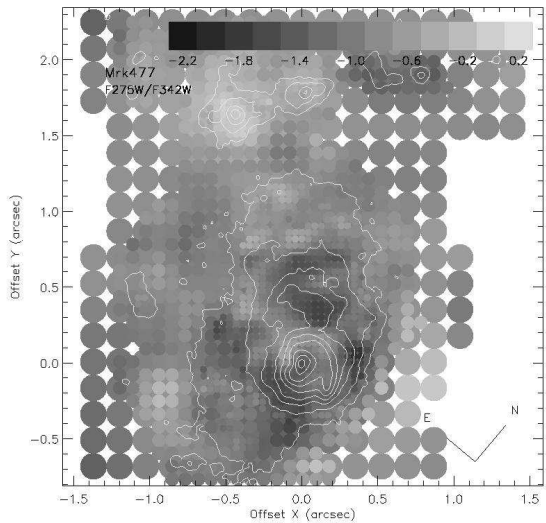}
\figcaption{The color map of the total flux, constructed from the
images of the two filters F342W and F275W. The total flux ratio is
converted to the spectral index $\alpha$ ($f_{\nu} \propto
\nu^{\alpha}$), and corrected for the small Galactic reddening of
$E(B-V) = 0.011$. The images were smoothed with a gaussian of FWHM 24,
12, and 6 pixels, and the color was calculated with 12, 6, and 3 pixel
bins, respectively. The smaller bins are overlaid on the larger bins.
Note that the binned pixels are squares, but presented in circles to
make the distinctions between small and large bins clear.
\label{fig_tfcolor_28}}
\end{figure}

On the other hand, we can obtain a color map of total flux with much
higher S/N, which is shown in Figure \ref{fig_tfcolor_28}.  The ratio
of the total flux in the F275W filter to that in the F342W filter has
been converted to the spectral index $\alpha$ (corrected for the small
Galactic reddening). The color map is a composite of three different
bins with three different smoothing sizes. We have convolved each
image with a gaussian of FWHM 24, 12, 6 pixels and generated the color
map with 12, 6, 3 pixel bins, respectively, and stacked these three
into one plot. For each bin case, the regions with the formal
1-$\sigma$ uncertainty (calculated using the counts in the smoothed
images) in the spectral index smaller than 0.5 are shown. The actual
uncertainty is estimated to be less than $\sim$ 0.25, by binning the
images with the smoothing FWHM size. For the error calculation, we
have assumed a 5\% uncertainty in the estimation of the background
counts and added this in quadrature to the statistical error.

\begin{figure}
\plotone{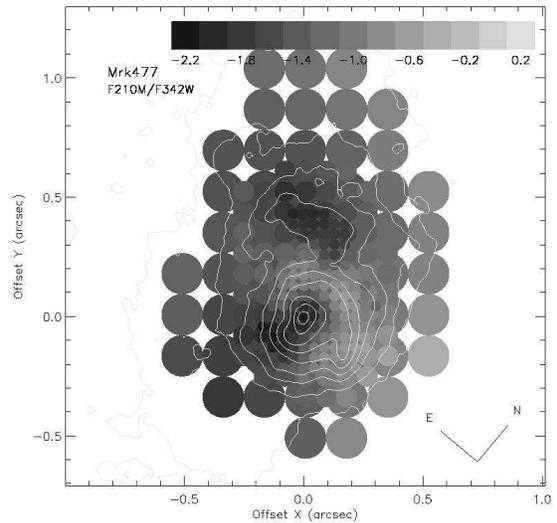}
\figcaption{The color map of the total flux, constructed from the
images of the two filters F342W and F210M. 
The presentation method is the same as Fig.\ref{fig_tfcolor_28}, but
the spatial window is smaller.
\label{fig_tfcolor_22}}
\end{figure}

This total flux color map, however, would have some uncertainty from
the emission line contamination as described in
\S\ref{sec-obs}. Therefore we have constructed another total flux
color map using our F342W image and the FOC/F210M image ($\lambda \sim
2180$\AA, taken in July 1995, two years before our observation;
Heckman et al. 1997), which is shown in Figure \ref{fig_tfcolor_22}
(note the spatial scale difference in Figures \ref{fig_tfcolor_28} and
\ref{fig_tfcolor_22}).  The bandpass of the F210M filter does not
contain any strong emission lines, and the larger wavelength interval
between the F210M and F342W filters results in a more accurate
measurement of the shape of the continuum without a strong influence
from the emission line contamination in the F342W filter (5\%
uncertainty in flux in each of the bands corresponds to $\Delta \alpha
\sim 0.4$ in F275W/F342W, but $\Delta \alpha \sim 0.2$ in
F210M/F342W).

In the several components in the central region (hotspot, fuzz, arm,
NE mirror), the F210M/F342W color map shows that the arm has a bluer
color ($\alpha \sim -0.8$), while the color is redder in the NE mirror
($\alpha \sim -1.7$) and the hotspot and fuzz ($\alpha \sim -2.0$),
and the red region extends to the south of the hotspot+fuzz. For these
components, roughly the same color distribution is also seen in the
F275W/F342W color map, but this map has other structures, which might
be from the emission line contamination. A knot structure seen outside
these nuclear components about $1.''7$ NE from the hotspot (see
Fig.\ref{fig_tfcolor_28}), which would possibly be a star-forming
region, have a bluer color ($\alpha \sim 0.0$ in F275W/F342W) than
that of the arm.

\section{Discussion}\label{sec-disc}

\subsection{The PA pattern in the inner region}

The central region of our imaging polarimetry data is contaminated by
the instrumental effect from the large intensity gradient around the
point-like source, which we call a hotspot. We have attempted to
subtract this effect using the PSFs through each polarizer. The data
after the subtraction might suggest that in addition to the NE mirror
region, the PA pattern in the nuclear vicinity of $\sim 0.''2$ radius
is also centrosymmetric.  This PA pattern appears to be surrounding
the hotspot region with a wide angle, as large as $\sim 180$\degr.

The uncertainty from the subtraction process, however, is large,
especially because the instrumental polarized flux in the PSF wing
would depend on the telescope focus/breathing at the time of the
observation.  Also, since the instrumental PA pattern around a point
source is roughly radial, an over-subtraction of this PA pattern from
the surrounding region would result in a perpendicular,
centrosymmetric (artificial) PA pattern, if this region is just
unpolarized.  The argument against this false artifact in our case of
Mrk 477, although not a strong argument, would be that in the N and NE
side of the hotspot the PA seems to rotate by the subtraction to
become closer to centrosymmetric, and in the NW, W, and SW side, the
pattern is already close to centrosymmetric before the subtraction
(see F275 image, Fig.\ref{fig_f275_papf}).

This possible centrosymmetric PA pattern extending $\sim 180$\degr\ at
the nuclear vicinity might be argued to suggest that the opening angle
of the radiation from the hidden nucleus is also 180\degr. However, we
definitely need more supporting evidence, such as an [OIII]/H$\alpha$
ratio map, that the high ionization region is really extended to this
wide angle.  Note that in the radio map (Heckman et al. 1997), where
three linearly aligned knots suggest a jet structure, the jet axis (PA
$\sim$ 30\degr) is along the direction to the NE mirror and not
perpendicular to the possible $\sim$ 180\degr\ opening, the axis of
which is at PA $\sim$ 150\degr.  Also the polarized light at the N,
NW, and W side of the hotspot, cospatial with the spiral arm
structure, might be simply due to the scattering of the bright hotspot
(and fuzz) light (not the hidden nuclear light), with the possible
high optical thickness at the spiral arm.

On the other hand, having all these uncertainties in mind, if the
opening angle of nuclear ionizing radiation projected to the sky plane
were indeed 180\degr, this would suggest that our line of sight is
marginally inside the conical radiation, and our line of sight might
be grazing the matter obscuring the nucleus.  If the obscuring matter
is in the torus-like geometry, our line of sight would be grazing the
surface of the torus. However, if this is the case, the scattered
light from the far side of the cone should also have a 180\degr\
opening angle (therefore, there should actually be no ``opening
angle''), though it is possibly weaker due to some absorption since it
is on the far side. If the scatterers are dust grains, their forward
scattering property would also contribute to the weakness of the
scattered light on the far side.  In our polarization image, the
polarized flux at the far side is not clear.

\subsection{The location of the hidden nucleus}\label{sec-disc-nucpos}

The location of the nucleus can be determined as the symmetric center
of the observed centrosymmetric PA pattern.  This can be calculated by
implementing a least-square fit of a simple centrosymmetric model to
the observed PA distribution with given errors for each data point
(see Kishimoto 1999 for the detailed method).  The symmetric center is
determined as a minimum \chisq\ point with an error circle defined by
a certain confidence level.

The result is shown in Figure \ref{fig_f275_papf}. We have used only
the data in the NE mirror region (with statistical S/N in $P$ larger
than 3), since the nuclear region is affected by the instrumental
effect.  For the error calculation, we assumed a 5\% uncertainty in
the background count estimation and added it to the statistical error
in quadrature.  Of the two plus signs in Figure \ref{fig_f275_papf},
the one at the right side which is almost at the hotspot is from the
F275W data, and the contour is its error circle of 99\% confidence
level.  The reduced \chisq\ was 0.96 with d.o.f. = 7. However, with
the F342W data, the reduced \chisq\ was found to be rather large, 1.7
with d.o.f. = 11, with the minimum \chisq\ point being $0.''06$ SE of
the hotspot (the other plus sign in Fig.\ref{fig_f275_papf}).  If we
include other systematic error, as described in Kishimoto (1999), the
reduced \chisq\ can go down to or less than 1. In that case, the 99\%
confidence level error circle extends to almost the same size as
that from the F275W data but to slightly different direction, and the
hotspot is at the edge of the error circle. This systematic difference
between the F275W and F342W data might be due to the instrumental
polarized flux as discussed in the previous section.  In terms of the
relation of the hidden nucleus location to the hotspot, our data do
not provide a strong upper limit on the distance between the nucleus
and the hotspot. However, our data are apparently consistent with the
two locations being coincident.

\subsection{Hotspot nature}
 
What is the nature of this hotspot radiation~? At least some portion
is expected to be scattered light, since the radiation seems to be at
least slightly polarized\footnotemark[2].  The low polarization does
not necessarily mean that the fraction of scattered light is
low. There might be efficient geometrical cancellation of the
polarization, or the scattered light might be of very small scattering
angle.  However, we argue below that scattered light does not seem to
be the primary component.

\footnotetext[2]{There might be some contribution from the
interstellar polarization intrinsic to Mrk 477, but this would
probably be small. A rather conservative upper limit would be $\sim$
3\% using the relation $P_{\rm max}$ (\%) $< 9 E(B-V)$ (Serkowski,
Mathewson, \& Ford 1975), based on the upper limit on the reddening of
the hotspot (see footnote 3). The Galactic interstellar polarization
is estimated to be very small from the polarization of nearby stars
($P < 0.1$\%; Tran 1995).}

If the hotspot is dominated by scattered light, its optical light
should not exceed the total optical scattered light, for which we can
set an upper limit from spectropolarimetry results.  If the fraction
of the scattered light is substantial, we should see, in the total
flux spectrum, the wings of the broad \ha\ and \hb\ lines, which have
been detected in the polarized flux (Tran 1995).  However, using the
spectropolarimetry data of Tran (1995; after the subtraction of an old
stellar population), we found little or no evidence of the \hb\ wing,
by overlaying various forbidden lines on to the \hb\ line.  From this,
we obtained a rather conservative upper limit on the fraction of the
scattered light to be 10\% (this corresponds to the intrinsic $P$ of
the scattered light to be $\gtrsim 10$\%, since the observed $P$ for
the continuum is about 1\%).  For the \ha\ region, essentially the
same limit was obtained from the red side of the \ha+[NII] lines
(though in the blue side there seem to be some more residuals).  This
upper limit is calculated to be $\sim 7 \times 10^{-17}$ \flam, by
adopting the total flux of $\sim 7 \times 10^{-16}$ \flam\ at
$4800$\AA\ (excluding an old stellar population; Tran 1995, Heckman et
al. 1997).  Now, this upper limit should be compared with the optical
flux of the hotspot, but unfortunately there is not an adequate HST
optical continuum image: the existing archival F606W images taken by
WFPC2 (on PC chip; $\lambda \sim$ 6000\AA) are saturated at the
hotspot, and also have a significant emission-line contribution of
$\sim$ 50\% (estimated using the spectrum of Tran 1995).  We obtain a
conservative lower limit of the optical hotspot continuum to be $\sim
3 \times 10^{-17}$ \flam, by implementing a formal synthetic
photometry with a 0.8 pixel radius aperture on the hotspot,
corresponding to the 2.5 pixel radius aperture used below for the FOC
images, and assuming a 50\% emission line contribution.  This would
not be too constraining. However, we can compare the above upper limit
on the total optical scattered light with the hotspot fluxes at other
wavelengths.

\begin{figure}
\plotone{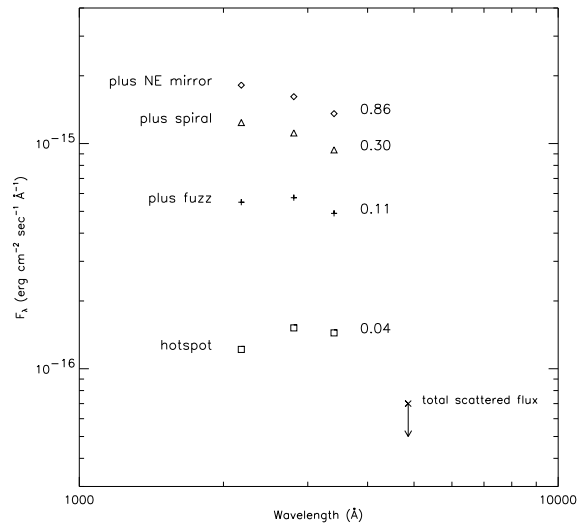}
\figcaption{The results of the synthetic aperture photometry with
various aperture size at different wavelength (F210M, F275W, and F342W 
filters). The aperture radii are indicated in arcseconds, and the
regions included in each aperture is also indicated. The cross at
4800\AA\ is the upper limit for the total amount of scattered light.
The fluxes have been corrected for the Galactic reddening of $E(B-V) =
0.011$. 
\label{fig_sed}}
\end{figure}

Using our two broad-band images and the FOC/F210M image ($\lambda \sim
2180$\AA; see previous section; Heckman et al. 1997), we have
constructed the UV/optical spectra of the hotspot and surrounding
regions, by implementing synthetic aperture photometry with several
aperture sizes. This is shown in Figure \ref{fig_sed}. The fluxes have
been corrected for Galactic reddening (see \S\ref{sec-intro}). The
smallest aperture (2.5 pixel radius) is essentially for the hotspot.
If the hotspot is dominated by scattered light, the above upper limit
on the total scattered light, indicated with a cross, gives a rather
blue color for the hotspot between 3400\AA\ and 4800\AA, $\beta <
-2.0$ where $F_{\lambda} \propto \lambda^{\beta}$ (or $\alpha > 0$
where $F_{\nu} \propto \nu^{\alpha}$).  The scattered light from the
hotspot would be much less than this upper limit, since this is the
upper limit for the sum of the scattered light. In our polarization
image with the F342W filter, the polarized flux from the central
$0.''14 \times 0.''14$ bin is roughly comparable to that detected in
the NE mirror. Based on this, if we assume that half of the total
scattered light at 4800\AA\ is from the hotspot, the limit of the
color becomes $\beta < - 4.0$ ($\alpha > +2.0$). Therefore, although
we cannot exclude the possibility that the hotspot is dominated by
scattered light, these colors seem to suggest the existence of another
significant component in addition to the scattered light.  Note that
the observed optical polarized flux is not too blue, $\beta = -1.6$
(or $\alpha = -0.4$; Tran 1995). The ratio of the polarized flux in
the $0.''14$ diameter aperture in our two images formally gives $\beta
= -0.6$ ($\alpha = -1.4$), though this is subject to several
uncertainties as discussed in \S\ref{sec-res}.

We do not expect that the hotspot is a heavily absorbed Seyfert 1
nucleus, if the size of the nucleus is smaller than the broad-line
region as usually thought.  We clearly do not see any strong direct
broad lines in the total flux, so the direct broad lines, if there are
any, must be heavily absorbed.  (For instance, conservatively, even if
the unabsorbed direct broad components have peak flux only comparable
to the narrow components, it would have to be suppressed at least by a
factor of $\sim 50$, based on the above upper limit on the broad line
component; this absorption corresponds to roughly $A_V \sim 4$.)  The
direct Seyfert 1 continuum would be absorbed even more.  This would be
inconsistent with the UV spectrum of the hotspot shown in Figure
\ref{fig_sed} which is not too red, unless $A_V/E(B-V)$ is extremely
high (note that the intrinsic color of direct continuum should be no
bluer than quasars, i.e. $\alpha \lesssim 0$ or $\beta \gtrsim -2$
[e.g. Neugebauer et al.  1987], and possibly redder than
this)\footnotemark[3].  Thus, conversely, in order to have enough
continuum light without having strong broad lines, we need another
continuum source outside the broad-line region, or the nuclear
continuum source has to be larger than the broad-line region.

\footnotetext[3]{An upper limit on the reddening of the hotspot can
roughly be obtained based on the hotspot spectrum shown in Figure
\ref{fig_sed}: even if the intrinsic (unabsorbed) spectral shape is as
blue as $\alpha \sim$ 0 (e.g. Neugebauer et al. 1989 for quasars;
Leitherer et al. 1999 for starbursts), the reddening $E(B-V)$ is
estimated to be $\sim$ 0.3, using Milky Way or SMC curves (Cardelli et
al. 1989; Witt \& Gordon 2000) with $A_V /E(B-V) = 3.1$.}

Heckman et al. (1997) showed that the far-UV ($1200-1600$\AA)
radiation from the central region is dominated by a starburst, based
on the presence of stellar wind features observed in the HST/GHRS
spectrum with a $1.''74 \times 1.''74$ aperture.  The radiation from
this central $1.''7$ region is essentially from the hotspot + fuzz +
arm + NE mirror, but the hotspot is only a minor part of the
radiation, as seen in Figure \ref{fig_sed} : $\sim 10$\% at 2800\AA\
and 3400\AA, and less at 2180\AA.  Therefore, for the hotspot nature,
this spectroscopic result does not provide a useful limit.  As
described in \S\ref{sec-res} and shown in Figure \ref{fig_tfcolor_22},
the color in the spiral arm is bluer than in the hotspot or the NE
mirror, so the spiral arm will be the primary flux contributor in the
central few arcseconds at shorter wavelength.  Therefore the stellar
wind features may well be originating primarily from this nuclear
spiral arm.  Note that Heckman et al also states that the scattered
light is a very minor part of the far-UV flux (less than 10\%), while
our data suggest that the flux from the NE mirror could be dominated
by scattered light. These are not inconsistent, since the NE mirror
flux is also a minor part of the whole integrated flux ($\sim 16$\%
within the $1.''7$ diameter in our two filters).

The variability of the UV/optical continuum reported for this galaxy
might be due to the hotspot, if the hotspot is the region really close
to the nucleus.  Note, however, that at least at the time of our
observation, the hotspot radiation was only $\sim$ 10\% of the UV
radiation in the central $\sim 2''$ diameter region as described
above.  Therefore, it would have to be variable by an order of
magnitude to cause the variability of a factor of $\sim 2$ in a large
aperture, which is reported by De Robertis (1987) and Kinney et
al. (1991), and this continuum brightening would have to occur without
a significant increase of the broad lines.  If this is the case, the
hotspot might be some part of the continuum source itself, which,
however, implies that the continuum source is extended over the
broad-line region.  The nature of the hotspot radiation would be of
great interest in the future high spatial resolution spectroscopy such
as by HST/STIS.

\section{Conclusions}\label{sec-conc}

We have presented HST UV imaging polarimetry data of the Seyfert 2
galaxy Mrk 477. For this galaxy, there are claims of variability in
the UV/optical, unusual for a Seyfert 2 galaxy.  It has a UV bright
pointlike hotspot in the central region, which is also peculiar among
nearby Seyfert 2 galaxies. Our data identify an off-nuclear scattering
region $\sim 0.''6$ ($\sim 500$ pc) NE from the hotspot. The data,
after the subtraction of the instrumental effect from the bright
hotspot region, might indicate that the scattered light is also
detected in the nuclear vicinity ($\sim 0.''2$ radius) and is extended
widely with a full opening angle of $\sim 180$\degr\ around the
hotspot region. This could lead us to consider the possibility that
our line of sight is grazing the matter obscuring the nucleus, which
might be the cause of some of the peculiar properties of this galaxy.
However, the uncertainty from the subtraction process is large and we
need more evidence to support this claim.

The hotspot location is consistent with the symmetry center of the PA
pattern, which represents the location of the hidden nucleus, but our
data do not provide a strong upper limit on the distance between the
hotspot and the symmetry center.  The hotspot radiation seems to be
slightly polarized, but it does not appear to be dominated by
scattered light.  Since we essentially do not see direct broad lines
and the UV spectral shape of the hotspot is not too red, we do not
expect the hotspot to be a heavily absorbed nucleus which is usually
thought to be inside the broad-line region.  There would be another
continuum source outside the broad-line region, or the nuclear
continuum source has to be larger than the broad-line region.

% ----- ----- ----- acknowledgments, appendix, references ----- -----

\acknowledgments

Support for this work was provided by NASA through grant number
GO-6702 to L. Kay from the Space Telescope Science Institute, which is
operated by AURA, Inc., under NASA contract NAS5-26555.  This work is
based on observations with the NASA/ESA Hubble Space Telescope,
obtained at the Space Telescope Science Institute.  The authors thank
the referee for carefully reading the manuscript and for helpful
comments.  This research has made use of the NASA/IPAC Extragalactic
Database (NED) which is operated by the Jet Propulsion Laboratory,
California Institute of Technology, under contract with the National
Aeronautics and Space Administration. M.K. was a Guest User, Canadian
Astronomy Data Centre, which is operated by the Herzberg Institute of
Astrophysics, National Research Council of Canada.

%\appendix

% ----- ----- ----- figures ----- ----- ----- -----

%\clearpage

\begin{figure}
\plotone{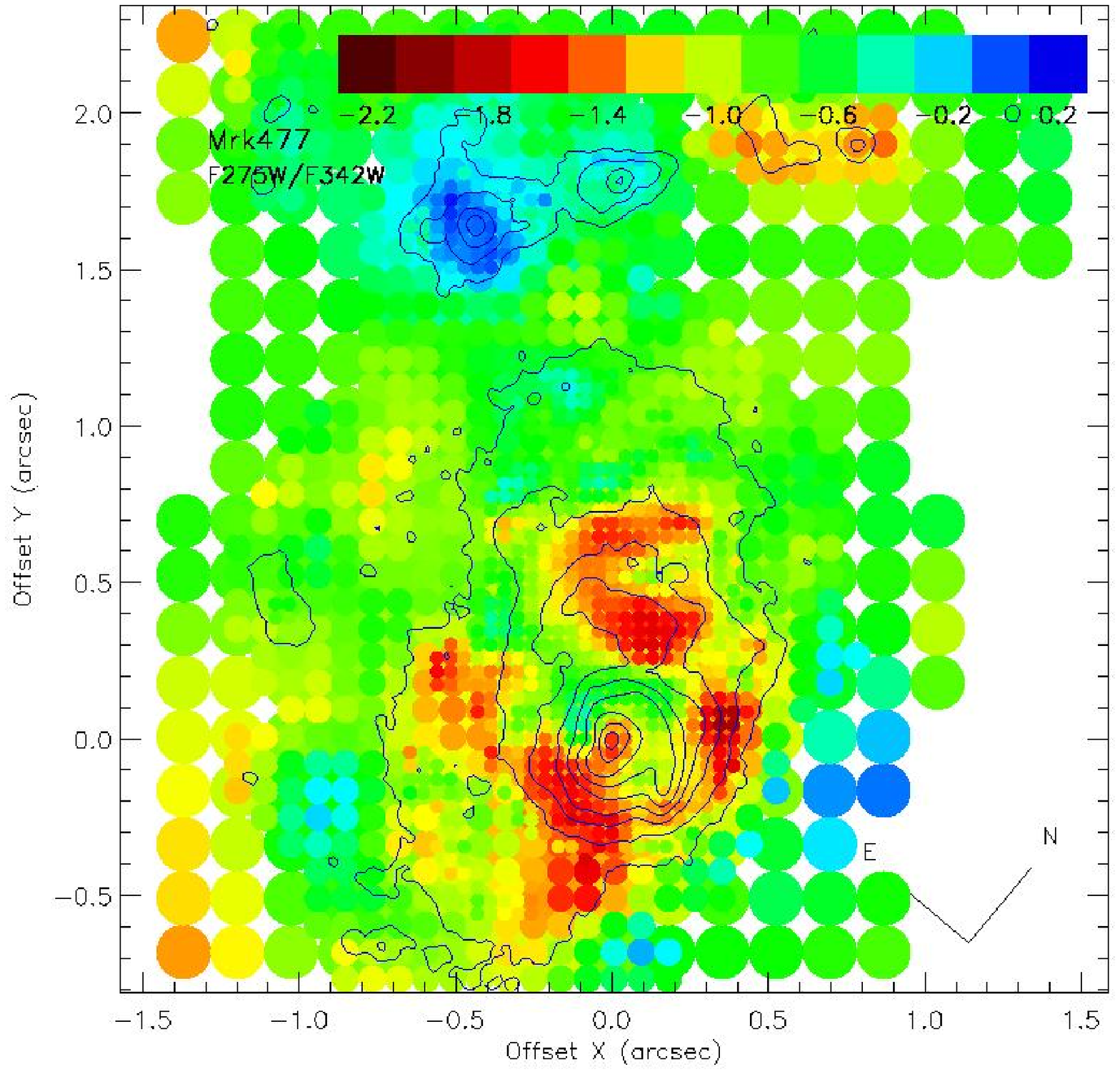}
\figcaption{Color version of Fig.\ref{fig_tfcolor_28}.}
\end{figure}

\begin{figure}
\plotone{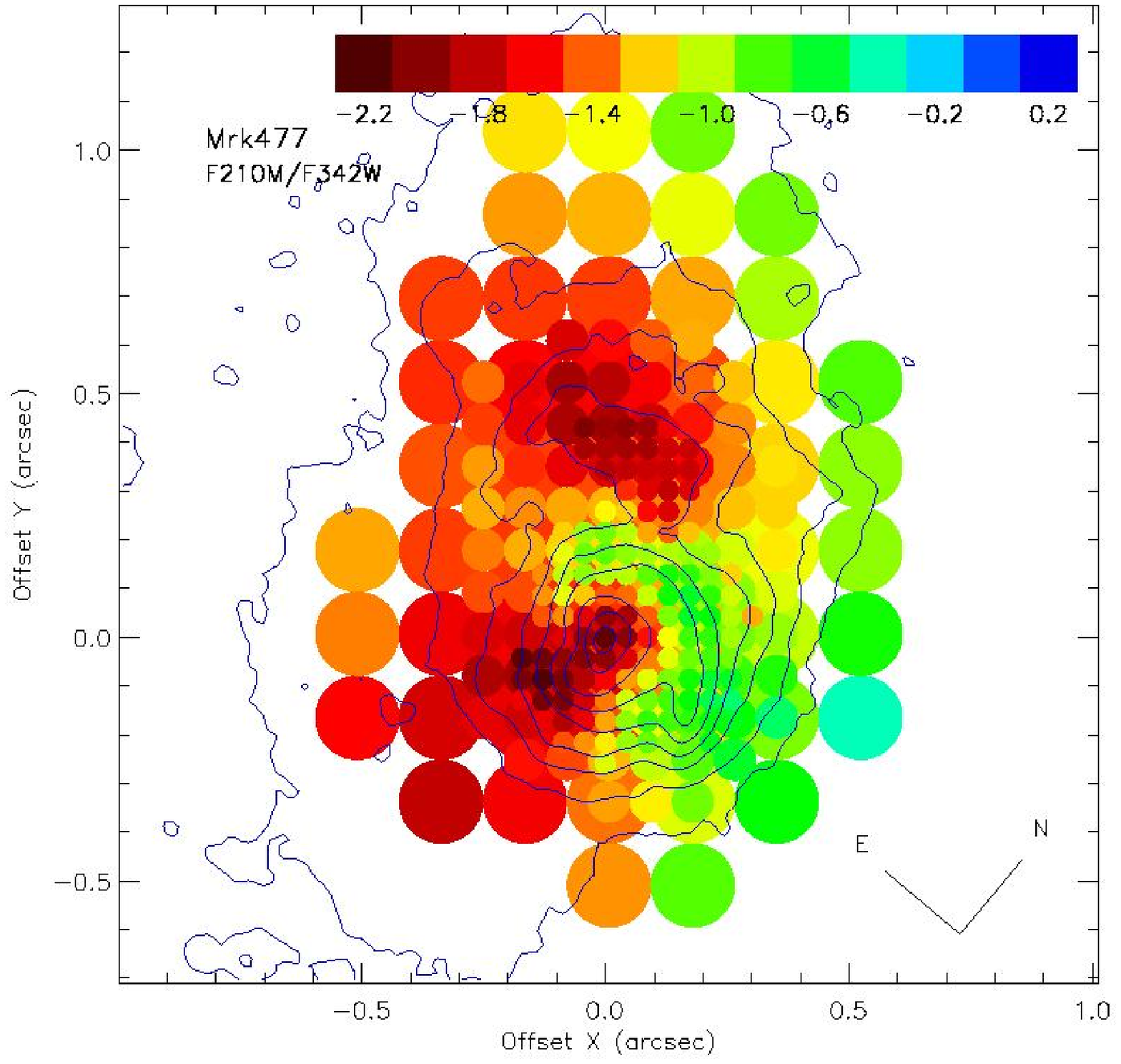}
\figcaption{Color version of Fig.\ref{fig_tfcolor_22}.}
\end{figure}

% ----- ----- ----- tables ----- ----- ----- -----

\clearpage

\begin{deluxetable}{cclccrrrrrrr}
%\rotate
%\tabletypesize{\scriptsize}
\tablecaption{Obtained FOC Data \label{tab-data}}
%\tablecolumns{6}
\tablewidth{0pt}
\tablehead{
\colhead{Rootname} & \colhead{Obs Date} &
\colhead{Filter} & \colhead{Exp Time (sec)} 
}
\startdata

x3md0201r, x3md0204r  & Aug 28, 1997 & F275W+POL0   & $2772 + 3063$&\\
x3md0207r, x3md020ar  & Aug 28, 1997 & F275W+POL60  & $3063 + 3063$&\\
x3md020dr, x3md020gr  & Aug 28, 1997 & F275W+POL120 & $3063 + 3063$&\\
x3md0301r  & Aug 30, 1997 & F342W+POL0   & $2772$&\\
x3md0304r  & Aug 30, 1997 & F342W+POL60  & $3063$&\\
x3md0307r  & Aug 30, 1997 & F342W+POL120 & $3063$&\\
x2rn0401t  & Jul 21, 1995 & F210M        & $2771.5$&\\
\enddata
\end{deluxetable}

\begin{deluxetable}{lcccclcccc}
%\rotate
%\footnotesize
\tabletypesize{\scriptsize}
\tablecaption{Comparison of polarization measurements\label{tab-comp}}
\tablewidth{0pt}
\tablecolumns{10}
\tablehead{
\colhead{} & \multicolumn{4}{c}{2800\AA} & \colhead{} &
\multicolumn{4}{c}{3400\AA}\\
\cline{2-5} \cline{7-10} \\
\colhead{instrument/aperture} &
\colhead{$F_{\lambda}$} & \colhead{$P F_{\lambda}$} & 
\colhead{$P$(\%)} & \colhead{P.A.(deg)} & \colhead{} &
\colhead{$F_{\lambda}$} & \colhead{$P F_{\lambda}$} & 
\colhead{$P$(\%)} & \colhead{P.A.(deg)}
}

\startdata
FOC  NE mirror\tablenotemark{a} & 
$2.66\pm0.02$ & $0.12\pm0.02$ & $4.3\pm0.8$ & $142\pm5$ &&
$2.283\pm0.006$ & $0.14\pm0.01$ & $6.2\pm0.4$ & $125\pm2$ \\

ground-based  $2''$ slit\tablenotemark{b} &
\nodata & \nodata & \nodata & \nodata &&
14.3 & $0.26\pm0.04$ & $1.8\pm0.3$ & $100\pm5$ \\

FOC  $2''$ diameter &
\nodata & \nodata & \nodata & \nodata &&
%$17.1\pm0.1$ & $0.20\pm0.13$ & $1.2\pm0.8$ & $185\pm17$  &&
$14.31\pm0.03$ & $0.31\pm0.04$ & $2.1\pm0.3$ & $94\pm4$ \\

FOS  $4.''3 \times 1.''4$\tablenotemark{c} &
$23.75\pm0.07$ & $0.53\pm0.13$ & $2.2\pm0.6$ & $132\pm7$ &&
\nodata & \nodata & \nodata & \nodata \\

FOC  $4.''3 \times 1.''4$\tablenotemark{d} &
$17.7\pm0.2$ & $0.29\pm0.25$ & $1.6\pm1.4$ & $187\pm23$  &&
\nodata & \nodata & \nodata & \nodata \\

\enddata

\tablecomments{$F_{\lambda}$ and $P F_{\lambda}$ are in units of
$10^{-16}$ \flam, and corrected for the small Galactic reddening
$E(B-V) = 0.011$. For the error estimation, we assumed a 5\%
uncertainty in the background counts estimation and added it in
quadrature to the statistical error. For the FOC data, the quoted
values for 2800\AA\ and 3400\AA\ are in the F275W and F342W filters,
respectively.}

\tablenotetext{a}{Synthetic aperture of $0.''8 \times 0.''6$ centered
at $(0.''00, 0.''55)$ in Fig.\ref{fig_f342_papf}.}

\tablenotetext{b}{Used the data of Kay (1994), integrated over
$3200-3600$\AA.} 

\tablenotetext{c}{Used the data of Cohen et al. (2002), integrated
over $2500-3100$\AA.  The minor axis of the observing aperture was at
PA=$-124$\degr. Pre-COSTAR data, while our FOC data is post-COSTAR.}

\tablenotetext{d}{Synthetic aperture simulating the FOS measurement,
which includes the whole central region and misses only a minor
portion of the NE mirror.}

\end{deluxetable}


\begin{thebibliography}{}

\bibitem[Antonucci \& Miller(1985)]{AM85} Antonucci, R., \& Miller,
J. S. 1985, \apj, 297, 621

\bibitem[Cardelli, Clayton, \& Mathis(1989)]{CCM89} Cardelli, J. A.,
Clayton, G. C., \& Mathis, J. S. 1989, \apj, 345, 245

\bibitem[Cohen et al.(2001)]{C01} Cohen et al. 2002, in preparation

\bibitem[De Robertis(1987)]{De87} De Robertis, M. M. 1987, \apj, 316,
597

\bibitem[Heckman et al.(1997)]{He97} Heckman, T. et al. 1997, \apj,
482, 114 

\bibitem[Hodge(1995)]{Ho95} Hodge, P. E. 1995, FOC Instrument Science
Report, 89 

\bibitem[Kay(1994)]{Ka94} Kay, L. E. 1994, \apj, 430, 196

\bibitem[Kinney et al.(1991)]{Ki91} Kinney, A. L., Antonucci,
R. R. J., Ward, M. J., Wilson, A. S., \& Whittle, M. 1991, \apj, 377,
100 

\bibitem[Kishimoto(1999)]{Ki99} Kishimoto, M. 1999, \apj, 518, 676

\bibitem[Kishimoto et al.(2001)]{Ki01} Kishimoto, M., Kay, L. E.,
Antonucci, R., Hurt, T. W., Cohen, R. D., Krolik, J. H. 2002, \apj,
in press

\bibitem[Leitherer et al.(1999)]{Le99} Leitherer, C. et al. 1999, \apjs,
123, 3

\bibitem[Nelson et al.(1996)]{Ne96} Nelson, C. H., MacKenty, J. W.,
Simkin, S. M., \& Griffiths, R. E. 1996, \apj, 466, 713

\bibitem[Neugebauer et al.(1987)]{Ne87} Neugebauer, G., Green, R. F.,
Matthews, K., Schmidt, M., Soifer, B. T., \& Bennett, J. 1987, \apjs,
63, 615

\bibitem[Malkan et al.(1998)]{Ma98} Malkan, M. A., Gorjian, V., \& Tam,
R. 1998, \apjs, 117, 25

\bibitem[Schlegel et al.(1998)]{Sc98} Schlegel, D. J., Finkbeiner,
D. P., \& Davis, M. 1998, \apj, 500, 525

\bibitem[Serkowski, Mathewson, \& Ford(1975)]{Se75} Serkowski, K.,
Mathewson, D. S., \& Ford, V. L. 1975, \apj, 196, 261

\bibitem[Simmons \& Stewart(1985)]{SS85} Simmons, J. F. L., \&
Stewart, B. G. 1985, \aa, 142, 100

\bibitem[Tran(1995)]{Tr95} Tran, H. D. 1995, \apj, 440, 565

\bibitem[Tran et al.(1992)]{Tr92} Tran, H. D., Miller, J. S., \& Kay,
L. E. 1992, \apj, 397, 452

\bibitem[Veilleux(1988)]{Ve88} Veilleux, S. 1988, \aj, 95, 1695

\bibitem[Witt \& Gordon(2000)]{WG00} Witt, A. N., \& Gordon,
K. D. 2000 \apj, 528, 799

\end{thebibliography}
\end{document}